\documentclass[aps,pra,twocolumn,amsmath,amssymb,nofootinbib,showpacs,superscriptaddress]{revtex4-1}
\usepackage[english]{babel}
\usepackage{latexsym}
\usepackage{graphics}
\usepackage{graphicx}
\usepackage{epsfig}
\usepackage{color}
\usepackage{bm}
\usepackage{amsmath}
\usepackage{amssymb}
\usepackage{amsthm}
\usepackage{dcolumn}
\usepackage{bm}
\usepackage{float}
\usepackage{hyperref}
\usepackage{color}
\usepackage{epstopdf}
\usepackage{cleveref, braket, comment, multirow}
\usepackage[svgnames]{xcolor}
\hypersetup{hidelinks,colorlinks=true,allcolors=DarkBlue}

\newcommand{\BLUE}[1]{{\color{Black} #1}}

\theoremstyle{remark}

\newcommand{\iswap}{\ensuremath{\mathtt{iSWAP}}}
\newcommand{\cnz}{\ensuremath{\mathtt{C}^{N-1}\mathtt{Z}}~}
\newcommand{\cnx}{\ensuremath{\mathtt{C}^{N-1}\mathtt{X}}~}
\newcommand{\cz}{\ensuremath{\mathtt{CZ}}~}
\newcommand{\cx}{\ensuremath{\mathtt{CX}}~}
\newcommand{\uab}{\ensuremath{U_{i\rightarrow j}}~}

\begin{document}

\preprint{APS/123-QED}

\title{Decomposing the generalized Toffoli gate with qutrits}

\author{A.S. Nikolaeva}
\affiliation{Russian Quantum Center, Skolkovo, Moscow 143025, Russia}
\affiliation{Moscow Institute of Physics and Technology, Moscow Region 141700, Russia}
\affiliation{National University of Science and Technology ``MISIS'', Moscow 119049, Russia}

\author{E.O. Kiktenko}
\affiliation{Russian Quantum Center, Skolkovo, Moscow 143025, Russia}
\affiliation{Moscow Institute of Physics and Technology, Moscow Region 141700, Russia}
\affiliation{National University of Science and Technology ``MISIS'', Moscow 119049, Russia}

\author{A.K. Fedorov}
\affiliation{Russian Quantum Center, Skolkovo, Moscow 143025, Russia}
\affiliation{National University of Science and Technology ``MISIS'', Moscow 119049, Russia}

\date{\today}
\begin{abstract}
The problem of finding efficient decompositions of multi-qubit gates is of importance for quantum computing, especially, in application to existing noisy intermediate-scale quantum devices, whose resources are substantially limited.
Here we propose a decomposition scheme for a generalized $N$-qubit Toffoli gate with the use of $2N-3$ two-qutrit gates for arbitrary connectivity. 
The fixed number of the required additional levels (the choice of qutrits is optimal) and the use of the iSWAP gate as a native operation make our approach directly applicable for ongoing experiments with superconducting quantum processors. 
Specifically, we present a blueprint of the realization of the proposed scheme for the Aspen-9 processor supporting quantum operations with qutrits.  
\end{abstract}

\maketitle

\section{Introduction}
A digital model of quantum computing can be considered as a natural extension of the classical computing model, which operates with quantum analogs of information bits known as qubits. 
During last decades, remarkable progress in the development of quantum information processing devices on various physical platforms has been performed~\cite{Martinis2019,Pan2020,Blatt2018-2,Wright2019,Browaeys2021,Lukin2021}. 
A peculiar feature of most existing physical systems used for quantum computing is the ability to operate in the larger states spaces, which are formed, for example, 
by accessible additional energy levels of ions or (artificial) atoms~\cite{Ruben2018,Zeilinger2018,Sanders2020}.
The use of such additional degrees of freedom is the basic idea of qudit-based quantum information processing. 
In the case of quantum computing, the use of qudits may help to reduce resources required for implementing quantum algorithms.
This is possible, first, by decomposing qudits on a set of two-level systems (qubits).
This approach has been widely studied both theoretically 
and experimentally~\cite{Farhi1998,Kessel1999,Kessel2000,Kessel2002,Muthukrishnan2000,Nielsen2002,Berry2002,Klimov2003,Bagan2003,Vlasov2003,Clark2004,Leary2006,Ralph2007,White2008,Ionicioiu2009,Ivanov2012,Li2013,Kiktenko2015,Kiktenko2015-2, Song2016,Frydryszak2017,Bocharov2017,Gokhale2019,Pan2019,Low2020,Jin2021,Martinis2009,White2009,Wallraff2012,Mischuck2012,Gustavsson2015,Martinis2014,Ustinov2015, Morandotti2017,Balestro2017,Low2020,Sawant2020,Pavlidis2021,Rambow2021}.
Despite its simplicity and efficiency, there is a number of limitations associated with finding optimal qudit-to-qubit mappings~\cite{Nikolaeva2021}. 
The second approach is to employ higher qudit levels can be used for substituting ancilla qubits, which is of specific interest in the problem of decomposing multi-qubit gates~\cite{Barenco1995}, 
such as the Toffoli gate~\cite{Ralph2007,White2009,Ionicioiu2009,Wallraff2012,Kwek2020,Baker2020,Kiktenko2020,Kwek2021,Galda2021,Gu2021}.
We note that in the pioneering experimental work on the realization of the Toffoli gate, the third energy level of the superconducting transmon qubit has been used~\cite{Wallraff2012}; 
this has been further exploited in Ref.~\cite{Galda2021} in experiments with fixed-frequency transmons on the IBM cloud-based superconducting processor and generalized~\cite{Gu2021}  for the $n$ qubits (qutrits) coupled within linear topology.
Systematic studies of this approach have demonstrated an ability to efficiently decompose multi-qubit gates: $2N-3$ qubit-qudit gates are required for decomposing the $N$-qubit generalized Toffoli gate~\cite{Kiktenko2020}.
However, this requires one to satisfy a general relation between the dimensionality of qudits and the topology of coupling map, specifically, 
for a given qudit one should have $d\geq{k+1}$, where $d$ is its dimension and $k$ is the number of its connections  to other qudits used within the decomposition~\cite{Kiktenko2020}.
The limited connectivity of existing noisy intermediate-scale quantum (NISQ) devices and sufficiently high level of errors in the operation with high energy levels poses a problem of relaxing conditions for the efficient implementation of multi-qubit gates.
Many works are then focused on revealing the potential of qutrits~\cite{Klimov2003,Bocharov2017,Gokhale2019,Galda2021}, 
which are possible to operate efficiently in experiments~\cite{White2008,Wallraff2012,Song2016,Jin2021,Galda2021,Wu2020,Blok2021,Hill2021,Ringbauer2021}.
We note that the use of qudits offers advantages also in quantum teleportation~\cite{Pan2019} and quantum communications~\cite{Gisin2002,Boyd2015}, 
as well as opens up opportunities for uncovering fundamental concepts of quantum mechanics~\cite{Li2013,Frydryszak2017,Zyczkowski2022}.

\begin{figure}
\center{\includegraphics[width=1\linewidth]{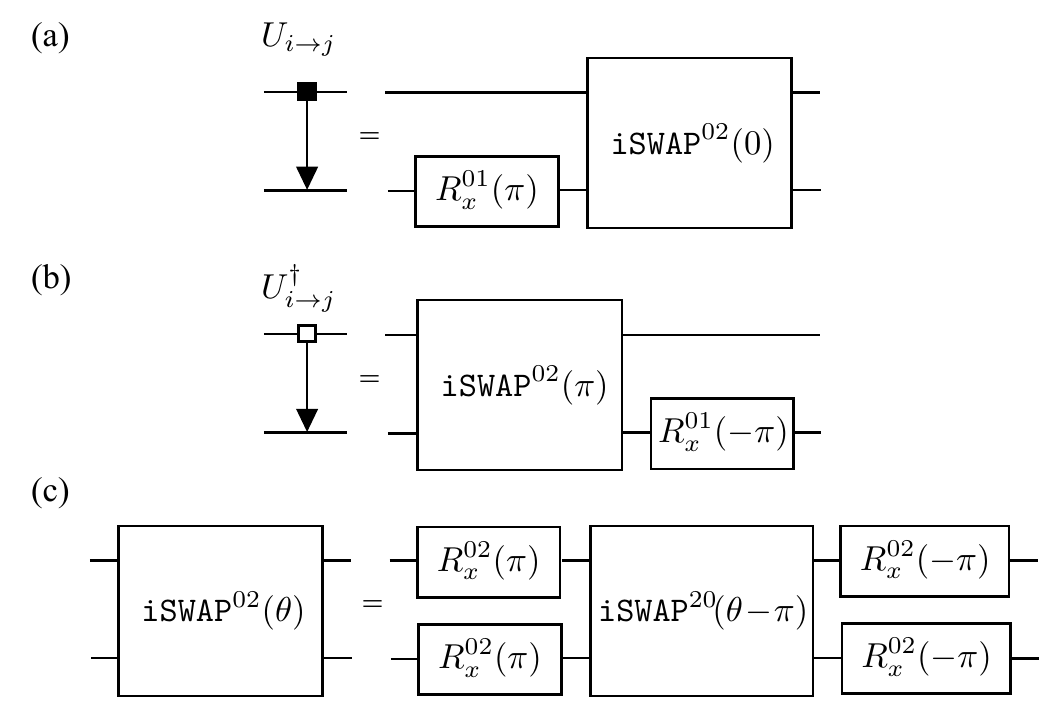}}
\caption{Decomposition of two-qutrit gate \uab (a) and its inverse $U_{i\rightarrow j}^\dagger$ (b) using $\mathtt{iSWAP^{02}}$ gate, which is native for superconducting qutrit-based platforms, is shown.
In (c) the transformation of $\mathtt{iSWAP^{02}}$ into  $\mathtt{iSWAP^{20}}$ using local operations is depicted.}
\label{fig::uab}
\end{figure}

\begin{figure*}[ht!]
	\includegraphics[width=1\linewidth]{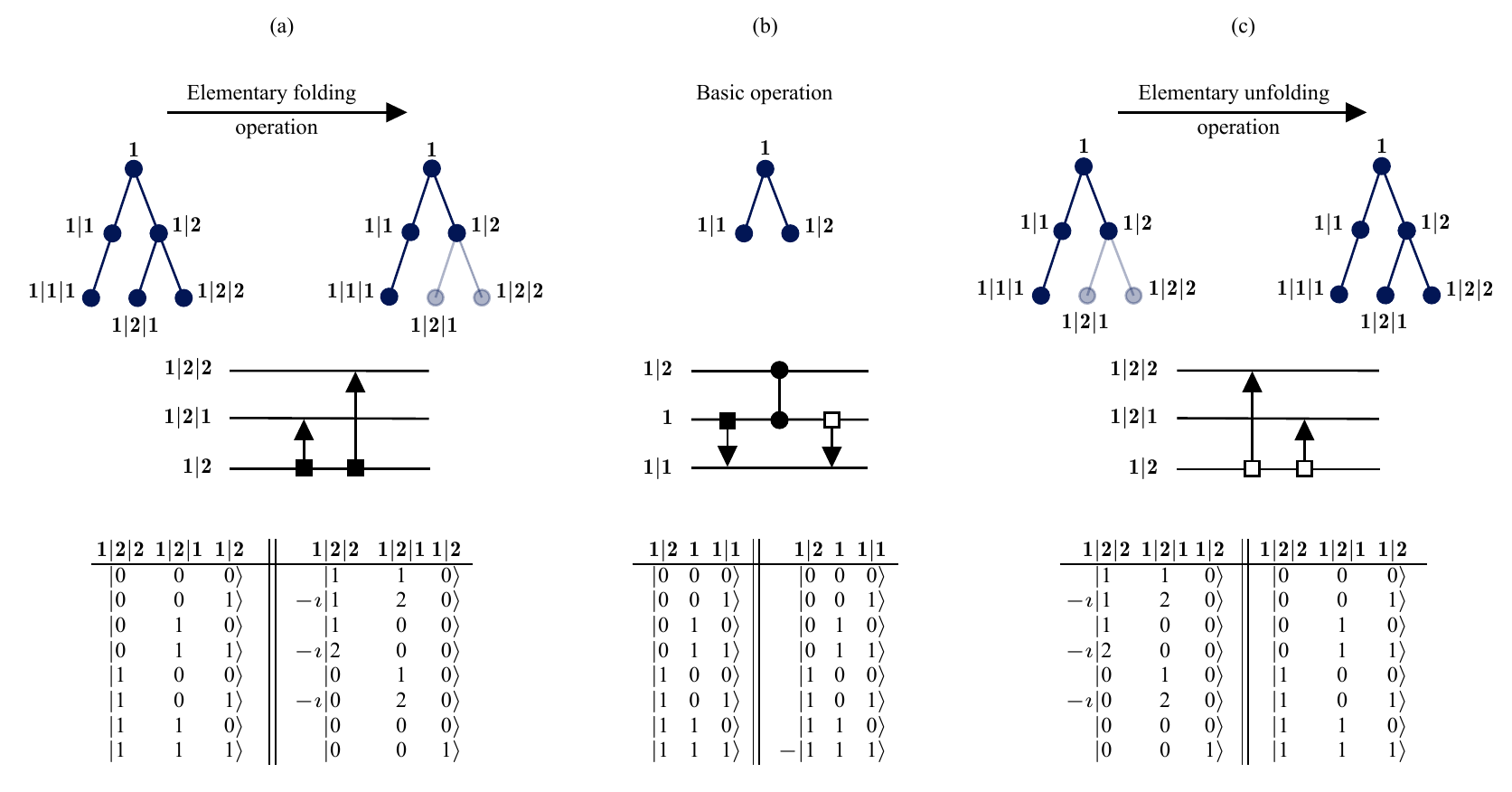}
	\vskip-3mm
	\caption{
        Three types of operations with the tree that correspond to $\cnz$ gate decomposition \BLUE{and evolution of the qutrits' states during these operations. 
        Before the decomposition, qutrits are initialized with the state of zero population on the ancillary level, therefore corresponding truth tables are presented for these qubit states of qutrits only.}
        In (a) the circuit structure for elementary folding operation is presented. It changes the tree structure by collapsing all siblings ${\bf s|1},\ldots, {\bf s}|n({\bf s})$ to their parent ${\bf s}$.
 	In (b) the circuit structure for the basic operation that applies a phase factor $-1$ to the input state if all qutrit of the level-one subtree are in the state $\ket{1}$ is shown. It does not change single-level tree structure.
        In (c) the circuit structure for elementary unfolding operation that serves as uncomputation of elementary folding operation is illustrated.
	}\label{fig::main}
\end{figure*}

In the present work, we improve recent results by developing a scheme for decomposing $N$-qubit generalized Toffoli gate using $2N-3$ two-qutrit gates by employing qutrits $(d=3)$ {\it only}; for summary, see Tab.~\ref{tab1}.
The fixed number of the required additional levels (the choice of qutrits is optimal) and the use of the \iswap~ gate as a native operation make our approach directly applicable for superconducting quantum processors, 
specifically, for the Aspen-9 processor supporting quantum operations with qutrits~\cite{Hill2021}.  
Although, we are focused here on the superconducting quantum computing setup, our approach in principle can be adopted for other physical platforms,
in particular, for trapped ions, for whose an ability to realize a multi-qudit processor has been demonstrated~\cite{Ringbauer2021}.

Our work is organized as follows.
In Sec.~\ref{sec:qutrit-proc}, we introduce main concepts behind the idea of a qutrit-based processor and describe the required quantum gates for its operation.
In Sec.~\ref{sec:Toffoli}, we describe the details of the proposed approach for decomposing the generalized Toffoli gate with qutrits.
In Sec.~\ref{sec:Aspen}, we discuss the application to existing architecture of superconducting qutirt-based processor.
We summarize our results in Sec.~\ref{sec:Conclusion}.

\section{Qutrit-based processor and gates}\label{sec:qutrit-proc}

We consider a system of $N$ qutrits, which are composed by three levels $\ket{0}$, $\ket{1}$, and $\ket{2}$, labeled by indices $i \in \{1,\dots,N\}$.
The first two levels $\ket{0}_i, \ket{1}_i$ of $i$th qutrit are considered as a qubit, whereas the third level $\ket{2}_i$ is as an ancillary level.
We focus on the problem of the $N$-qubit gate decomposition, and then assume that the initial and the final states of the $N$-qutrit system can be considered an $N$-qubit states, i.e., population of the ancillary states $\ket{2}_{i}$ should be zero. 

As basic single-qutrit operations, we employ the following gates:
\begin{equation}
	R_{x(y)}^{01}(\varphi)=e^{-\imath \lambda_{1(2)} \varphi}, \quad
	R_{x(y)}^{12}(\varphi)=e^{-\imath \lambda_{6(7)} \varphi}.
\end{equation}
Here standard notations of Gell-Mann matrices are used:
\begin{equation}
	\begin{aligned}
		\lambda_{1}&={\begin{pmatrix}0&1&0\\1&0&0\\0&0&0\end{pmatrix}}, \quad 
		\lambda_{2}&={\begin{pmatrix}0&-\imath&0\\\imath&0&0\\0&0&0\end{pmatrix}},\\
		\lambda_{6}&={\begin{pmatrix}0&0&0\\0&0&1\\0&1&0\end{pmatrix}}, \quad
		\lambda_{7}&={\begin{pmatrix}0&0&0\\0&0&-\imath\\0&\imath&0\end{pmatrix}}.
	\end{aligned}
\end{equation}
We note that these operations correspond to experimentally accessible transitions in superconducting qutrit-based processors~\cite{Blok2021}.
The $0\to2$ transition, \BLUE{which cannot be accessed directly at least in certain experimental setups~\cite{Blok2021}, can be implemented via the sequence of} intermediate operations: 
\begin{equation}
	R_{x(y)}^{02}(\varphi)=e^{-\imath(\pi/2)\lambda_6}e^{-\imath(\varphi/2)\lambda_{1(2)}}e^{\imath(\pi/2)\lambda_6}.
\end{equation}
\BLUE{If the $0\to2$ transition is directly available, this provides the further simplification of the proposed scheme.}

As native two-qutrit operations, we consider $\mathtt{iSWAP^{02}(\theta)}$ and $\mathtt{iSWAP^{20}(\theta)}$ gates representing a `standard' $\mathtt{iSWAP}$ operation in the $|11\rangle$-$|02\rangle$ and $|11\rangle$-$|20\rangle$ subspaces, 
respectively, with an additional off-diagonal controllable phase \cite{Hill2021}:
\begin{equation}
    \begin{aligned}
    \iswap^{02}(\theta)\ket{11}&=-\imath e^{-\imath \theta}\ket{02},\\
    \iswap^{02}(\theta)\ket{02}&=-\imath e^{-\imath \theta}\ket{11},\\
    \iswap^{02}(\theta)\ket{xy}&=\ket{xy}, ~\text{for}~ xy \neq 11, 02,
    \end{aligned}
\end{equation}
and 
\begin{equation}
    \begin{aligned}
    \iswap^{20}(\theta)\ket{11}&=-\imath e^{-\imath \theta}\ket{20},\\
    \iswap^{20}(\theta)\ket{20}&=-\imath e^{-\imath \theta}\ket{11},\\
    \iswap^{20}(\theta)\ket{xy}&=\ket{xy}, ~\text{for}~ xy \neq 11, 20.
    \end{aligned}
\end{equation}
Such quantum gates are native for superconducting quantum computing platforms~\cite{Blok2021,Oliver2019,Hill2021}.

\begin{table*} 
\caption{
Comparison of qudit-based \cnz (\cnx) gate decompositions. 
$^{*}$The logarithmic scale of the decomposition depth is achieved for the optimal structure of the tree $\widetilde{E}$, that can be chosen i.e. within all-to-all coupling map.
$^{**}$Here $d_i$ is a dimension of a particular qudit, and $k_i$ is a number of other qudits connected to this qudit within decomposition.
$^{***}$Due to the specificity of the decomposition, the exact number of two-qudit gates is given for the case $N=2^\alpha$, $\alpha\geq 2$.
$^{****}$The number is given for the case of no measurement-based feed-forward operations.
}
\begin{ruledtabular}
\begin{tabular}{ccccccc}
&This work & Ref.~\cite{Gu2021} &Ref.~\cite{Gokhale2019}& Ref.~\cite{Kiktenko2020} & Refs.~\cite{White2009,Ralph2007} &Ref.~\cite{Ionicioiu2019}  \\ \hline
Depth & {$\bm{\mathcal{O}(\log N)}^*$ }& $\mathcal{O}(N)$ & $\bm{\mathcal{O}(\log N)}$ & $\bm{\mathcal{O}(\log N)}^*$& $\mathcal{O}(N)$  & $\mathcal{O}(N)$\\
\multirow{2}{*}{Qudit type} & \multirow{2}{*}{\bf qutrits} & \multirow{2}{*}{\bf qutrits} & \multirow{2}{*}{\bf qutrits} & qudits,  & single qudit, & single qudit, \\
&  & & & ${d_i\geq k_i+1}^{**}$  & $d=N+1$ & $d=N$ \\
\# of ancillas &{\bf0} &{\bf 0} &{\bf 0}&{\bf 0}&{\bf 0}& 1 \\
Coupling map & {\bf arbitrary} &linear &all-to-all & {\bf arbitrary} & star & star\\
\# of 2-body gates  & $\bm{2N-3}$ & $\bm{2N-3}$ &  ${6N-11}^{***}$  &$\bm{2N-3}$& $\bm{2N-3}$ &$\bm{2N-3}^{****}$\\
\end{tabular}
\end{ruledtabular}\label{tab1}
\end{table*}

To define a coupling map that determines the possibility of applying at least one of the described two-qutrit gates,
we introduce $E$ as a set of ordered pairs $(i, j)$, such that $i, j \in \{1,\dots, N\}, i < j$.
We then suppose that it is possible to execute $\mathtt{iSWAP^{02}(\theta)}$ or $\mathtt{iSWAP^{20}(\theta)}$ between qutrits $i$ and $j$ if and only if $(i,j)\in E$.
We also assume the graph corresponded to the coupling map is connected.

Combinations of native two-qutrit gates together with single-qutrit rotations give us more complex gates.
An illustrative example is an implementation of a standard controlled-phase gate $\cz$ that can be obtained as sequence of two \iswap$^{02}(0)$ (or \iswap$^{20}(0)$) gates:
the \cz gate then can be transformed into the \cx gate by surrounding \cz with $R_y^{01}(-\pi/2)$ and $R_y^{01}(\pi/2)$ rotations.

We also introduce an auxiliary gate $U_{i\rightarrow j}$ acting on $i$th and $j$th qudit obtained as a sequence of $R_x^{01}(\pi)$ and \iswap$^{02}(0)$ [see Fig.~\ref{fig::uab}(a)].
The idea behind of $U_{i\rightarrow j}$ is that it leaves $i$th qutrit in state $\ket{1}_{i}$ only if both qutrits $i$ and $j$ are in the state $\ket{1}$. 
Otherwise, the state of $i$th qutrit becomes $\ket{0}_{i}$.
The action of  $U_{i\rightarrow j}$ on qubit levels of input qutrits is as follows:
\begin{equation}
	\begin{aligned}
		U_{i\rightarrow j}\ket{00}_{ij}&=\ket{01}_{ij},\\
		U_{i\rightarrow j}\ket{01}_{ij}&=\ket{00}_{ij},\\
		U_{i\rightarrow j}\ket{10}_{ij}&=-\imath\ket{02}_{ij},\\
		U_{i\rightarrow j}\ket{11}_{ij}&=\ket{10}_{ij}.
	\end{aligned}
\end{equation}
We note that $U_{i\rightarrow j}$ can be also obtained from
\iswap$^{20}$ gate using \iswap$^{20}$  decomposition shown in Fig.~\ref{fig::uab}(c).

\section{Generalized Toffoli gate decomposition}\label{sec:Toffoli}

The generalized $N$-qubit Toffoli gate $\cnx$ inverts the state of the target qubit if and only if all $N-1$ control qubits are in the state $\ket{1}$, and acts as identity otherwise.
Its implementation can be reduced to surrounding the generalized $N$-qubit controlled-phase gate \cnz by $R_y(\pi/2)$ and $R_y(-\pi/2)$ on the target qubit.
We note that \cnz has a symmetrical form and is given by
\begin{equation}\label{eq:global_CZ_gate}
\begin{aligned}
	&\cnz\ket{1\ldots 1}_{1\ldots N} = -\ket{1\ldots 1}_{1\ldots N},\\
	&\cnz\ket{x_1\ldots x_N}_{1\ldots N} = \ket{x_1\ldots x_N}_{1\ldots N},
\end{aligned}
\end{equation}
for $\prod_i x_i\neq 1$.

To construct the \cnz gate decomposition, we first choose an acyclic connected graph (tree) $\widetilde{E}$ within the coupling graph $E$ by removing, if necessary, some edges from $E$.
All two-qutrit gates that we use in our decomposition are taken from $\widetilde{E}$ only.
We also choose the root of the tree.
As it is shown below, the optimal choice of $\widetilde{E}$ and the root has to provide the minimal height of tree, i.e. the minimal number of edges between the root an the furthest node. 

We relabel qutrits according to the choice of $\widetilde{E}$ and the root.
Each node (qutrit) is identified by its `address' in the tree.
We denote the root node as ${\bf 1}$.
The siblings of the node $({\bf s})$ are denoted by ${\bf s|1}, {\bf s|2}, \ldots {\bf s}|n({\bf s})$, where $n({\bf s})$ is a number of siblings of $({\bf s})$.

Our method for the \cnz~gate decomposition is along the lines of the approach of Ref.~\cite{Kiktenko2020}.
Specifically, we introduce three groups of operations with a tree $\widetilde{E}$, namely: (i) folding operation, (ii) basic operation, and (iii) unfolding operation [see Fig.~\ref{fig::main}(b)].
Each group of operations corresponds to adding gates to the decomposition circuit that is initialized as empty one.

We first apply the folding operation that transforms the original tree $\widetilde{E}$ to a one-level form in which only the root $\bf 1$ and its siblings ${\bf 1|1},\ldots,{\bf 1}|n({\bf 1})$ remain.  
To achieve this, we sequentially apply the sequence of so-called elementary folding operations [see Fig.~\ref{fig::main}(a)].
Each elementary folding operation collapses a set of leaves ${\bf s|1},\ldots,{\bf s}|n({\bf s})$  having the same parent ${\bf s}$.
At the same time, we add $U_{\bf s \rightarrow {\bf s}|\ell}$ gate for $\ell={\bf 1},\ldots, n({\bf s})$ to the decomposition circuit [see Fig.~\ref{fig::main}(a)].

Application of $U_{{\bf s} \rightarrow{\bf s}|\ell}$ to the qutrits ${\bf s}$ and ${\bf s}|\ell$ keeps ${\bf s}$ in the state $\ket{1}_{\bf s}$  if and only if ${\bf s}$ and ${\bf s}|\ell$ were initially in the state $\ket{11}_{{\bf s},{\bf s}|\ell}$.
Thus, after elementary folding operation on a subtree ${\bf s}$, ${\bf s|1},\ldots,{\bf s}|n({\bf s)}$,
qutrit  ${\bf s}$ is left in the state $\ket{1}_{\bf s}$ if and only if all qutrits ${\bf s}$, ${\bf s|1},\ldots,{\bf s}|n({\bf s)}$ were in the state $\ket{1}$ before its start.
Otherwise, the  ${\bf s}$ turns into the state 0 and some of leaves ${\bf s}|\ell$ turn into the ancillary state 2.
We note that after the whole folding procedure the root's siblings ${\bf 1|1,\ldots,1}|n({\bf 1})$ transform into the state $\ket{1\ldots1}_{{\bf 1|1,\ldots,1}|n({\bf 1})}$ if and only if all the qutrits except the root have been initialized in the state $\ket{1}$. 
\BLUE{The transformation of initial qubit states of the qutrits after application of the elementary folding operation is presented in the truth table in Fig.~\ref{fig::main}(a)}.

Further, we apply the basic operation that involves only qutrits that correspond to the to the root and its leaves [see Fig.~\ref{fig::main}(b)].
In analogy with the elementary folding operation, the basic operation consists of sequential application of $U_{\bf 1 \rightarrow 1|\ell}$ for $\ell={\bf 1},\ldots n({\bf 1})-1$. 
We then apply the \cz~gate to the root and the last leaf ${\bf 1}|n({\bf 1})$, and repeat the first $n({\bf 1})-1$ operations in reverse order replacing $U_{\bf 1 \rightarrow 1|\ell}$ by $U_{\bf 1 \rightarrow 1|\ell}^\dagger$.
The idea of the basic operation is to acquire a phase factor $-1$ if all qutrits of the level-one subtree are in the state $\ket{1}$. 

At the last step, we employ the unfolding operation that is a `mirror reflection' of the folding.
It consists of a sequence of elementary unfolding operations that return the the graph to its original form, and serves for an uncomputation purposes [see Fig. \ref{fig::main}(c)].
As the result, \BLUE{after the application of all three groups of operations,} the state taken from computational basis and belonging to the qubit subspace acquires the phase factor of $-1$ if all qutrits were in the state $\ket{1}$, 
and remains the same otherwise.

One can see that by the design of the scheme $2N-3$ two-qutrit gates are required: $2N-4$ $\iswap$ gates (or its reversals) and single \cz gate.
We note that (un)folding operations applied to siblings ${\bf{p}|1},\ldots,{\bf{p}}|n({\bf{p}})$ and ${\bf{q}|1},\ldots,{\bf{q}}|n({\bf{q}})$ of different parent nodes ${\bf{p}}$ and $\bf{q}$ can be performed simultaneously.
Therefore, some blocks of the decomposition circuit can be performed in parallel and the decomposition circuit depth depends on the tree height and its structure, more specifically on the number of siblings of each node. 
If the tree is a complete $\kappa$-ary tree, where $\kappa$ is fixed parameter, then circuit depth belongs to $\mathcal{O}(\log N)$. 

\begin{figure}[ht!]
\center{\includegraphics[width=0.675\linewidth]{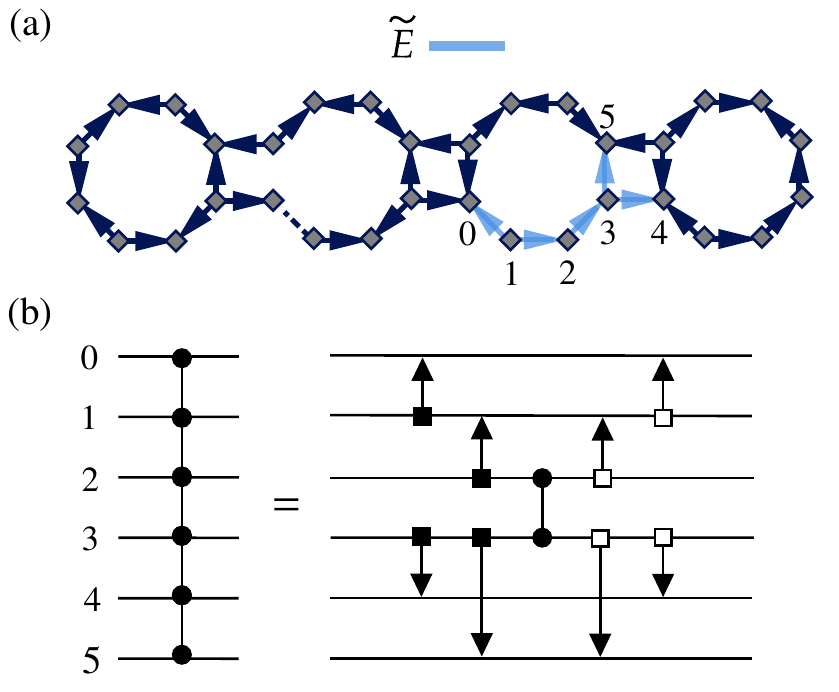}}
\caption{
Example of the $\mathtt{C}^{5}\mathtt{Z}$ gate decomposition for the existing qutrit-based superconducting quantum processor.
In (a) coupling map of the Aspen-9 architecture~\cite{Hill2021} is shown.
Each incoming arrow indicates a transmon that transitions to the second excited state $\ket{2}$ in the corresponding two-qutrit operation.
Subgraph $\widetilde{E}$ corresponds to the qutrits that are chosen for decomposition of $\mathtt{C}^{5}\mathtt{Z}$ gate. 
In (b) the resulting decomposition circuit for the $\mathtt{C}^{5}\mathtt{Z}$ gate is illustrated.}
\label{fig::aspen9}
\end{figure}

We provide the detailed comparison of our approach with other qudit-based \cnz (\cnx) gate decompositions in Tab. \ref{tab1}.
One can see that our approach combines the best features of previously proposed schemes in terms of circuit depth, number of ancillas, qudit type (our approach uses qutrits only), coupling map, and the number of required two-qutrit gates.
Our approach allows decomposing the $N$-qubit \cnz (\cnx) gate with $2N-3$ two-qutrit gates with arbitrary coupling map.
We also note that in comparison with the method of Ref.~\cite{Kiktenko2020}, our decomposition requires less time for the execution.
This is because we use the \iswap~ gate as the main operation, whereas the approach introduced in Ref.~\cite{Kiktenko2020} uses the \cz gate; the realization of the \cz gates requires us to implement the \iswap~ gate twice.

\section{Application to existing architecture of superconducting qutirt-based processor}\label{sec:Aspen}

We use our method for the decomposition of the $\mathtt{C}^{5}\mathtt{Z}$ gate for the Aspen-9 processor containing 32 qutrits with a honeycomb-like coupling map [see Fig. \ref{fig::aspen9}a].
We note that calibration of this processor is organized in such a way that the for each pair of coupled qutrits, only one type of interaction ($\iswap^{02}$ or $\iswap^{20}$) is chosen. 

Let us consider 6-qubit $\mathtt{C}^{5}\mathtt{Z}$ decomposition on qutrits with coupling map, which is illustrated in Fig. \ref{fig::aspen9}a. 
On the subgraph $\tilde{E}$~ 6-qubit $\mathtt{C}^{5}\mathtt{Z}$ is the implemented with 9 two-qutrit gates, 8 \iswap~ and 1 \cz gate [see Fig. \ref{fig::aspen9}b].
We note that types of \iswap~gates depends on the chosen nodes/qutrits in $\tilde{E}$. 
If in $\tilde{E}$ a parent node ${\bf s}$ has an incoming arrow from its sibling ${\bf s}|\ell$, then $\uab$ gate must be implemented with $\iswap_{20}$ in its core. 
This result is directly applicable for the decomposition of multi-qubit gates of the Aspen-9 processor. 

\section{Conclusion}\label{sec:Conclusion}

We have developed a scheme for the generalized $N$-qubit Toffoli gate, which is efficient in terms of the required number of $2N-3$ two-qutrit gates, types of qudits (qutrits only), and coupling between information carriers. 
The developed approach can combined with other techniques allowing one to reduce the complexity of the implementation of quantum algorithms by using qudits~\cite{Nikolaeva2021}.
We have used native gates of the superconducting quantum processors for our scheme, which makes it directly applicable for such category of NISQ devices.
We also expect that approach can be extended to the use for other platforms, for example, the trapped ion platform, in which qudits can be efficiently controlled~\cite{Ringbauer2021,Low2020}, and molecules~\cite{Sawant2020,Ruben2018}. 

\section*{Acknowledgments} 

The work was supported by the Russian Science Foundation Grant No. 19-71-10091.

\bibliography{bibliography-qudits.bib}

\end{document}